%% file: main.tex
\documentclass[lettersize,journal]{IEEEtran}
\usepackage{amsmath,amsfonts}
\usepackage{algorithmic}
\usepackage{algorithm}
\usepackage{array}
\usepackage[caption=false,font=footnotesize]{subfig}
\usepackage{textcomp}
\usepackage{stfloats}
\usepackage{url}
\usepackage{verbatim}
\usepackage{graphicx}
\usepackage{cite}
\usepackage{amsfonts, amssymb, cuted}
\usepackage{mathtools}
\usepackage{xcolor}
\usepackage{soul}
\usepackage{nicefrac}
\usepackage{enumitem}
\usepackage{acro}
\usepackage{pifont}

\usepackage{tikz}
\usepackage{pgfplotstable}
\usepackage{pgfplots}
\usetikzlibrary{plotmarks}
  \usetikzlibrary{arrows.meta}
  \usepgfplotslibrary{patchplots}
  \usepackage{grffile}
  \pgfplotsset{plot coordinates/math parser=false}
  \newlength\figureheight
  \newlength\figurewidth
   \pgfplotsset{compat=1.11,
    /pgfplots/ybar legend/.style={
    /pgfplots/legend image code/.code={%
       \draw[##1,/tikz/.cd,yshift=-0.25em]
        (0cm,0cm) rectangle (3em,8pt);},
   },
}

\input{MyCommands}

\begin{document}

\title{A Novel Space-Time Coding Architecture for Rydberg Atomic Quantum Receiver-Based Systems}

\author{Asifa Zannat,~\IEEEmembership{Student Member,~IEEE,}
        Milad Abolpour,~\IEEEmembership{Member,~IEEE,}
        Dani Korpi,~\IEEEmembership{Member,~IEEE,}
        Mikko A. Uusitalo,~\IEEEmembership{Senior Member,~IEEE,}
        Mikko Valkama,~\IEEEmembership{Fellow,~IEEE,}
        and~Ertugrul Basar,~\IEEEmembership{Fellow,~IEEE}%
\thanks{A. Zannat, M. Abolpour, M. Valkama, and E. Basar are with the Department of Electrical Engineering, Tampere University, Tampere, Finland (e-mail: asifa.zannat@tuni.fi; milad.abolpour@tuni.fi; mikko.valkama@tuni.fi; ertugrul.basar@tuni.fi). D. Korpi and M. A. Uusitalo are with Nokia Bell Labs, Espoo, Finland (e-mail: dani.korpi@nokia-bell-labs.com; mikko.uusitalo@nokia-bell-labs.com).}
        }%

\maketitle

\begin{abstract}
Rydberg atomic quantum receivers (RAQRs) offer high sensitivity and wide tunability, but their magnitude-based readout yields a nonlinear model incompatible with conventional complex-valued multi-input multi-output (MIMO) processing. We propose a low-complexity space-time coding framework for point-to-point RAQR-assisted MIMO links. Data symbols are encoded using real orthogonal designs, while strong-reference heterodyne reception yields an equivalent real-valued linear model. The preserved orthogonality enables matched filter symbol-wise detection without matrix inversion or vector search. An analytical bit error probability expression is derived, proving the proposed scheme achieves the full transmit-receive diversity. Simulations validate the analysis and demonstrate improved performance over spatial multiplexing benchmarks.
\end{abstract}


\section{Introduction}
\label{sec:intro}

The stringent requirements of future wireless systems, including high sensitivity, broad frequency coverage, and compact receiver deployment, are challenging for conventional radio-frequency (RF) front-ends \cite{cui2026rare}. Classical RF receivers rely on antenna-induced currents followed by analog filtering, amplification, mixing, and calibration, which makes their sensitivity strongly affected by thermal noise, hardware impairments, and frequency-dependent front-end design constraints~\cite{gong2025raqr}. In contrast, Rydberg atomic quantum receivers (RAQRs) exploit the strong coupling between incident electromagnetic fields and highly excited Rydberg atoms, enabling RF-to-optical field transduction through atom-light interactions~\cite{zhang2024rydberg}. Owing to the large dipole moments and rich energy-level structure of Rydberg atoms, RAQRs can provide ultra-high field sensitivity, wide tunability from near-DC to THz frequencies, and wavelength-independent receiver dimensions~\cite{fancher2021rydberg}. RAQRs, however, do not directly recover both the amplitude and phase of the incident RF field, since conventional atomic readout primarily depends on the magnitude of the RF-induced Rabi frequency and is therefore insensitive to the RF phase~\cite{liu2025aware}. This limitation can be mitigated using heterodyne RAQR architectures, in which a reference RF field, or local oscillator (LO), is introduced into the atomic sensor to enable phase-sensitive detection and support phase-modulated symbol detection~\cite{simons2019rydberg}.

Motivated by these capabilities, recent studies have investigated RAQR-enabled wireless communication and sensing from several perspectives, including atomic multi-input multi-output (MIMO) reception~\cite{cui2025atomicmimo}, spatial multiplexing for multi-band operation~\cite{zhu2025raq}, multi-user transmission~\cite{gong2025multiuser}, MIMO precoding~\cite{cui2025mimo}, learning-based receiver and transceiver designs~\cite{song2025csi,kang2026deepq}, and statistical RF signal detection~\cite{atapattu2025detection,atapattu2026multi}. To handle the RAQR-specific observation model, iterative phase-retrieval-based algorithms have been proposed to recover transmitted signal vectors~\cite{cui2025atomicmimo}, while modulation-aware and CSI-free learning-based methods have been developed to reduce receiver-side modeling and estimation requirements~\cite{zhu2025multiuser,song2025csi}. Other approaches modify the transmitter or receiver structure to obtain a more tractable signal model. In particular, phase-rotated symbol spreading (PRSS) transmits each complex MIMO signal vector together with its phase-rotated replica over two time slots, allowing the receiver to combine two real-valued RAQR measurements and reconstruct an effective complex-valued linear MIMO observation~\cite{liu2026prss}. This reconstructed model allows standard MIMO detectors to be applied to the received symbol vectors. However, PRSS primarily focuses on realizing a spatial-multiplexing model, whereas the design of RAQR-specific transmission schemes that explicitly exploit transmit diversity and enable low-complexity symbol-wise detection remains largely unexplored.

In this letter, we propose a low-complexity space-time coding architecture for a point-to-point (P2P) MIMO link, where a conventional RF transmitter equipped with $N_{\rm t}$ antennas communicates with an RAQR consisting of $N_{\rm r}$ vapor cells as receive elements. Inspired by the orthogonal space-time block coding (OSTBC) framework in~\cite{tarokh1999space}, the transmitter encodes the information symbols into a space-time codeword tailored to the real-valued RAQR observation model. At the receiver, a heterodyne RAQR architecture with a strong reference field is used to obtain phase-sensitive measurements. The resulting equivalent real-valued system preserves the orthogonality of the proposed codeword, allowing the received signals to be linearly combined and the transmitted real symbols to be detected independently with low complexity. We further derive the analytical bit error probability (BEP) of the proposed scheme and show that it achieves a diversity order of $N_{\rm t}N_{\rm r}$, demonstrating that the proposed scheme provides superior bit error rate (BER) performance compared with spatial-multiplexing-based baselines.

\textit{Notation:} Bold lowercase and uppercase letters denote vectors and matrices, respectively.  For a vector $\Bv$ and matrix $\BM$, $[\Bv]_i$, $[\BM]_{i,j}$, $[\BM]_{i,:}$, and $[\BM]_{:,i}$ denote the $i$-th entry, $(i,j)$-th entry, $i$-th row, and $i$-th column, respectively.  Moreover, $[\BM]_{a:b,:}$ and $[\BM]_{:,a:b}$ denote the submatrices formed by rows $a$ to $b$ and columns $a$ to $b$ of $\BM$, respectively.  $\BI_N$ denotes the identity matrix of size $N$.  The operators $(\cdot)^{\rm T}$, $\|\cdot\|_{\rm F}$, $\operatorname{diag}(\cdot)$, $\mathbb{E}[\cdot]$, $\operatorname{col}\{\cdot\}$, $\Re\{\cdot\}$, $\Im\{\cdot\}$, and $\angle(\cdot)$ denote transpose, Frobenius norm, diagonalization, expectation, column-wise stacking, real part, imaginary part, and phase, respectively.  Finally, $[n]=\{1,\ldots,n\}$, and $\mathcal{CN}(\cdot,\cdot)$ and $\mathcal{N}(\cdot,\cdot)$ denote circularly symmetric complex Gaussian and real Gaussian distributions, respectively.

\section{System Model}
\label{sec:system_model}
\begin{figure}[t]
    \centering
    \includegraphics[width = \columnwidth]{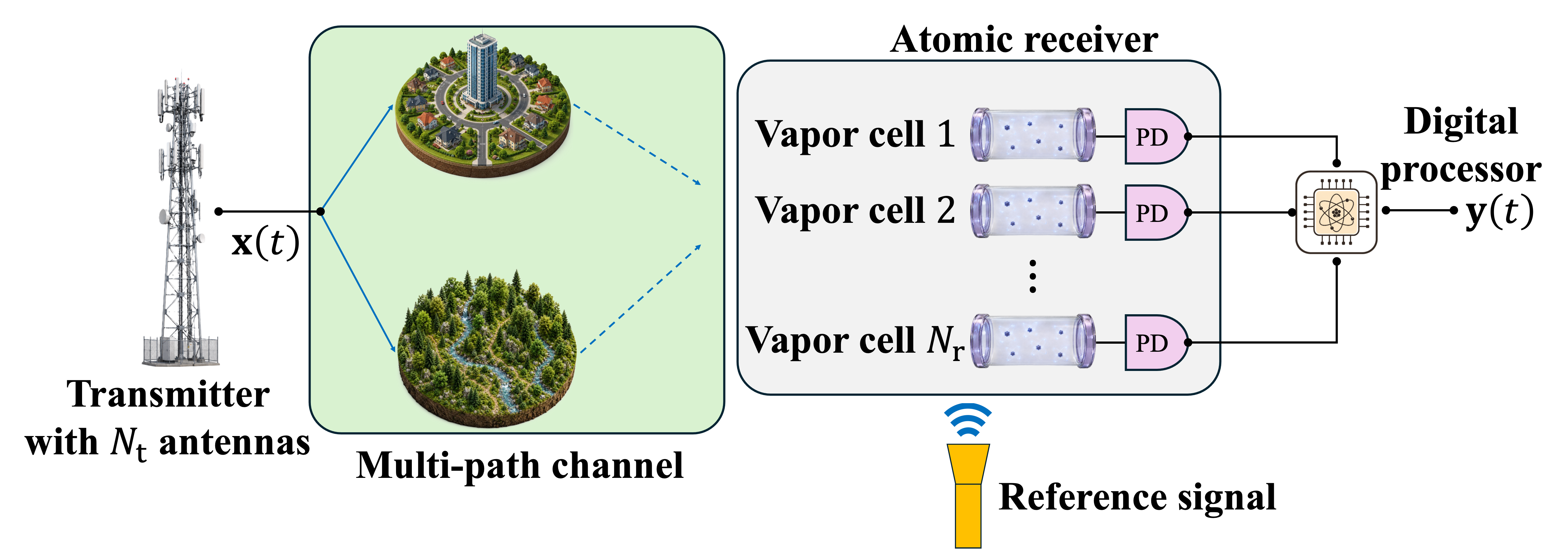}
    \caption{System model of the considered RAQR-assisted MIMO link with a receiver-side reference signal.}
    \label{fig:placeholder}
\end{figure}

We consider a P2P MIMO link where a conventional RF transmitter equipped with $N_{\rm t}$ antennas communicates with a Rydberg atomic quantum receiver equipped with $N_{\rm r}$ receive elements.  As depicted in Fig.~\ref{fig:placeholder}, each receive element is modeled as a vapor-cell-based atomic sensor that converts the incident RF electric field into an optical readout. Unlike conventional coherent RF receivers~\cite{gong2025raqr}, an RAQR directly measures a magnitude-dependent atomic response and therefore does not inherently preserve the phase of the received signal. To enable coherent baseband processing and retrieve the phase-related information, we assume that a known reference RF signal is injected at the receiver side~\cite{simons2019rydberg}. In this regard, let $\Bx(t)\in\mathbb{C}^{N_{\rm t} \times 1}$ denote the transmitted baseband signal vector during channel use $t$, and let $\Bb \in\mathbb{C}^{N_{\rm r} \times 1}$ denote the reference signal observed at the RAQR array, such that the $m$-th entry of the reference vector is modeled as $[\Bb]_m = \frac{\boldsymbol{\mu}^{\rm T}\boldsymbol{\epsilon}_{{\rm b}m}}{\hbar} g_{{\rm b}m} e^{j\phi_{{\rm b}m}}$, where $\boldsymbol{\epsilon}_{{\rm b}m}$, $g_{{\rm b}m}$, and $\phi_{{\rm b}m}$ denote the polarization vector, channel gain, and phase shift of the reference signal to the $m$-th receive element, respectively, $\boldsymbol{\mu}$ is the transition dipole-moment vector, and $\hbar$ is the reduced Planck constant~\cite{cui2025atomicmimo}.  We note that the reference signal $\Bb$ is provided by a receiver-side LO and is constant. Hence, the received signal during the $t$-th channel use, denoted by $\By(t)$, is modeled as
\begin{equation}
\label{eq:nonlinear_raqr_model}
\By(t)
=
\left|
\BH\Bx(t)+\Bb +\Bw(t)
\right|,
\end{equation}
where the absolute value is applied element-wise, $\Bw(t)\sim\mathcal{CN}(\mathbf{0},\sigma^2\BI_{N_{\rm r}})$ is the vector of complex additive noise samples, and $\BH \in \mathbb{C}^{N_{\rm r} \times N_{\rm t}}$ represents the effective channel matrix. Now, considering a multipath channel with $L$ propagation paths between each transmit and receive element, and following a similar concept to~\cite{cui2025atomicmimo},  the effective channel coefficient from the $k$-th transmit antenna to the $n$-th receive element (vapor cell)  is modeled as
\begin{equation}
\label{eq:multipath_channel}
[\BH]_{n,k}
=
\frac{1}{\sqrt{L}}
\sum_{\ell=1}^{L}
g_{nk\ell}
\frac{\boldsymbol{\mu}^{\rm T}\boldsymbol{\epsilon}_{nk\ell}}{\hbar}
e^{j\phi_{nk\ell}},
\end{equation}
where $g_{nk\ell}\sim\mathcal{CN}(0,1)$ denotes the complex small-scale fading coefficient of the $\ell$-th path, $\boldsymbol{\epsilon}_{nk\ell}$ is the polarization vector of the incident electric field, and $\phi_{nk\ell}$ is the propagation-induced phase shift. Here, the factor $1/\sqrt{L}$ normalizes the accumulated multipath contribution so that the average channel power does not scale with $L$. 

Under the strong-reference signal condition, i.e., $|[\Bb]_m|\gg |[\BH\Bx(t) + \Bw(t)]_m|$ for all $m$, employing the first-order approximation, and subtracting the known reference magnitude from $\By(t)$ in \eqref{eq:nonlinear_raqr_model}, the observed signal at the receiver is approximated as~\cite{cui2025mimo}: 
\begin{equation}
\label{eq:linearized_raqr_model}
\tilde{\By}(t)
=
\By(t)-|\Bb|
\approx
\Re\left\{
\operatorname{diag}\left(e^{-j\angle\Bb}\right)
\BH\Bx(t)
\right\}
+
\tilde{\Bw}(t),
\end{equation}
where $\tilde{\Bw}(t)$ denotes the resulting real-valued noise with $\tilde{\Bw}(t)\sim\mathcal{N}(\mathbf{0},\frac{\sigma^2}{2}\BI_{N_{\rm r}})$. Consequently, defining the phase-compensated effective channel as
\begin{equation}
\label{eq:effective_channel}
\BG
=
\operatorname{diag}\left(e^{-j\angle\Bb}\right)\BH,
\end{equation}
the equivalent baseband model in \eqref{eq:linearized_raqr_model} becomes
\begin{equation}
\label{eq:equivalent_raqr_model}
\tilde{\By}(t)
\approx
\Re\{\BG\Bx(t)\}
+
\tilde{\Bw}(t).
\end{equation}
Now, let $\BG_{\rm R}=\Re\{\BG\}$, $\BG_{\rm I}=\Im\{\BG\}$ and define the real-equivalent transmit vector as $\tilde{\Bx}(t)
=
\begin{bmatrix}
\Re\{\Bx(t)\}^{\rm T} &
\Im\{\Bx(t)\}^{\rm T}
\end{bmatrix}^{\rm T} \in\mathbb{R}^{2N_{\rm t}\times 1}$, then, \eqref{eq:equivalent_raqr_model} is written as
\begin{equation}
\label{eq:linear_system_modified}
\tilde{\By}(t)
\approx
\begin{bmatrix}
\BG_{\rm R} & -\BG_{\rm I}
\end{bmatrix}
\tilde{\Bx}(t)
+
\tilde{\Bw}(t).
\end{equation}
This real-valued linear model forms the basis for the OSTBC construction and detection framework developed in the sequel.

\subsection{Illustrative Example}

Before presenting the general construction, we first illustrate the proposed model for the case of two transmit antennas. Let $  \Bs  = [s_1,s_2,s_3,s_4]^{\rm T}  \in \mathbb{R}^{4 \times 1}$ denote the vector of real information symbols to be transmitted. For $N_{\rm t}=2$, the RAQR real-equivalent model contains $2N_{\rm t}=4$ real transmit dimensions, corresponding to the real and imaginary components of the two complex transmit signals. Hence, the proposed codeword can be constructed by starting from a real orthogonal design for four real transmit
dimensions and then mapping its rows to the real and imaginary parts of a complex two-antenna transmit block.
Specifically, for $N_{\rm t}=2$, the proposed OSTBC codeword spans four channel uses and is given by
\begin{equation*}
\label{eq:Nt2_codeword}
    \BX
    =  \frac{1}{2}
    \begin{bmatrix}
    s_1+js_3 & -s_2+js_4 & s_3-js_1 & s_4+js_2\\
    s_2+js_4 & s_1-js_3 & s_4-js_2 & -s_3-js_1
    \end{bmatrix}.
\end{equation*}
The rows of $\BX$ correspond to the two transmit antennas, while the columns correspond to four consecutive channel uses. For instance, during the channel use $t = 1$, the transmitted vector is given by $\Bx(1) = [\BX ]_{:,1} = \frac{1}{2} \begin{bmatrix}
    s_1+js_3\\
    s_2+js_4
    \end{bmatrix}$. Accordingly, during the four channel uses, the receiver observes the signal $\By(t)$ shown in \eqref{eq:nonlinear_raqr_model}. Hence, by employing \eqref{eq:effective_channel}, the phase-compensated RAQR channel is written as $ \BG =   \BG_{\rm R} + j\BG_{\rm I}$. Now, for $k = 1,2$, let  $\Bg_{{\rm R},k}$ and $\Bg_{{\rm I},k}$ be the $k$-th column of $\BG_{\rm R}$ and $\BG_{\rm I}$, respectively. By adopting the strong-reference signal assumption presented in \eqref{eq:equivalent_raqr_model}, stacking the RAQR outputs $\Tilde{\By}(t)$, with $t \in [4]$, as $ \tilde{\By}_{\rm eq}
    =
    \operatorname{col}
    \{
    \tilde{\By}(1),
    \tilde{\By}(2),
    \tilde{\By}(3),
    \tilde{\By}(4)
    \}
    \in
    \mathbb{R}^{4N_{\rm r} \times 1}$, and defining $\tilde{\Bw}_{\rm eq}
    =
    \operatorname{col}
    \{
    \tilde{\Bw}(1),
    \tilde{\Bw}(2),
    \tilde{\Bw}(3),
    \tilde{\Bw}(4)
    \}
    \in
    \mathbb{R}^{4N_{\rm r} \times 1}$, the received observations over the four channel uses can be given as follows:
\begin{equation*}
\label{eq:Nt2_stacked_model}
    \tilde{\By}_{\rm eq} = \BG_{\rm eq}\Bs + \tilde{\Bw}_{\rm eq},
\end{equation*}
where the equivalent channel matrix $\BG_{\rm eq}$ is given by
\begin{equation*}
\label{eq:Nt2_Geq}
    \BG_{\rm eq}
    = \frac{1}{2}
    \begin{bmatrix}
    \Bg_{{\rm R},1} & \Bg_{{\rm R},2} & -\Bg_{{\rm I},1} & -\Bg_{{\rm I},2}\\
    \Bg_{{\rm R},2} & -\Bg_{{\rm R},1} & \Bg_{{\rm I},2} & -\Bg_{{\rm I},1}\\
    \Bg_{{\rm I},1} & \Bg_{{\rm I},2} & \Bg_{{\rm R},1} & \Bg_{{\rm R},2}\\
    \Bg_{{\rm I},2} & -\Bg_{{\rm I},1} & -\Bg_{{\rm R},2} & \Bg_{{\rm R},1}
    \end{bmatrix}
    \in
    \mathbb{R}^{4N_{\rm r}\times 4}.
\end{equation*}
Here, it can be verified that $\BG_{\rm eq}^{\rm T}\BG_{\rm eq} =  \frac{1}{4} \|\BG\|_{\rm F}^{2}  \BI_4$, which enables the four real information symbols $s_1,s_2,s_3,$ and $s_4$ to be decoupled and independently detected by applying a matched filter (MF) combiner.

\subsection{OSTBC Construction}
In this section, we present the general structure to build the RAQR-OSTBC codewords. An OSTBC codeword is represented by a complex matrix $\BX\in\mathbb{C}^{N_{\rm t}\times T}$, where $T$ denotes the number of consecutive channel uses such that during the $t$-th channel use, the transmitter sends the $t$-th column of $\BX$, i.e., $\Bx(t) = [\BX]_{:,t}$. To construct $\BX$, we first form a real orthogonal design $\BC\in\mathbb{R}^{2N_{\rm t}\times T}$ by following a similar approach as in \cite{tarokh1999space}. Indeed, the $2N_{\rm t}$ rows of $\BC$ represent the real-equivalent transmit dimensions. Here, the aim is to design $\BC$ such that the vector of real information symbols $\Bs=[s_1,s_2,\ldots,s_T]^{\rm T}\in\mathbb{R}^{T \times 1}$ is detected by the receiver within $T$ consecutive channel uses, where each $s_i$ is drawn from an $M$-ary real constellation $\mathcal{S}$, achieving the spectral efficiency of $\eta = \log_{2}(M)$ bits per channel use (bpcu). To this end, $T$ is chosen to satisfy $\rho(T)\geq 2N_{\rm t}$,  where $\rho(T)$ denotes the Hurwitz--Radon function~\cite{geramita1979orthogonal}. In this sense, if $T=2^{4a+b}u$, where $u$ is odd and $b\in\{0,1,2,3\}$, then $\rho(T)=8a+2^b$. 

The construction of $\BC$ is based on a set of real matrices $\BA_1,\BA_2,\ldots,\BA_{2N_{\rm t}} \in \mathbb{R}^{T\times T}$, where these matrices are chosen to satisfy
\begin{equation}
\label{eq:A conditions}
        \begin{cases}
           \BA_i^{\rm T}\BA_i= \BI_{T} &  i=1,\ldots,2N_{\rm t} \\
           \BA_i^{\rm T}\BA_j + \BA_j^{\rm T}\BA_i = \mathbf{0} & i,j = 1, \ldots, 2N_{\rm t}, i \neq j
        \end{cases}.
\end{equation}
Indeed, the constraints in~\eqref{eq:A conditions} ensure the orthogonality of the resulting real-valued codeword $\BC$ and allow the real information symbols to be decoupled by linear processing at the RAQR. Consequently, the $k$-th row of $\BC$, with $k \in [2N_{\rm t}]$, is structured as $[\BC]_{k,:} = (\BA_{k}\Bs)^{\rm T}$, which results in $\BC\BC^{\rm T} = \|\Bs\|^2 \BI_{2N_{\rm t}}$, as per \eqref{eq:A conditions}. In the next step, we map the real-valued design $\BC$ to the codeword complex matrix $\BX$. To this end, matrix $\BC$ is decoupled into matrices $\BC_{\rm R}$ and $\BC_{\rm I}$, such that $\BC_{\rm R} = [\BC]_{1:N_{\rm t} , :}$ and $\BC_{\rm I} = [\BC]_{N_{\rm t}+1:2N_{\rm t} , :}$, i.e., $\BC = \begin{bmatrix}
    \BC_{\rm R}^{\rm T} &
    \BC_{\rm I}^{\rm T} 
\end{bmatrix}^{\rm T}$. Finally, the complex OSTBC transmit block is obtained as: 
\begin{equation}
\label{eq:X_def}
\BX = \frac{1}{\sqrt{2N_{\rm t}}}(\BC_{\rm R} + j\BC_{\rm I}),
\end{equation}
where the normalization factor $\frac{1}{\sqrt{2N_{\rm t}}}$ ensures that the average transmit energy per channel use does not increase with $N_{\rm t}$. In the following, we present the RAQR detection process for the OSTBC-based framework.

\subsection{Detection Process}
Here, we derive the low-complexity symbol-wise detection rule for the proposed OSTBC structure. Accordingly, we first reformulate the received signal during each channel use. As stated earlier, the transmit signal during the $t$-th channel use takes the form of  $\Bx(t) = [\BX]_{:,t}$. Now, by employing \eqref{eq:X_def} and considering  $[\BA_i\Bs]_t=[\BA_i]_{t,:}\Bs$, $\Bx(t)$ is rewritten as follows: 
\begin{equation}
\label{eq:x_t_from_C}
\Bx(t) = 
\frac{1}{\sqrt{2N_{\rm t}}}\Bigg( 
\begin{bmatrix}
[\BA_1]_{t,:}\\ 
\vdots\\
[\BA_{N_{\rm t}}]_{t,:}
\end{bmatrix} + j
\begin{bmatrix}
[\BA_{N_{\rm t}+1}]_{t,:}\\
\vdots\\
[\BA_{2N_{\rm t}}]_{t,:}
\end{bmatrix}
\Bigg) \Bs.
\end{equation}
Next, by substituting \eqref{eq:x_t_from_C} into \eqref{eq:equivalent_raqr_model},  the observed signal at the receiver during the $t$-th channel use becomes
\begin{equation}
\label{eq:y_t_detection}
\tilde{\By}(t) \approx \BG_t\Bs + \tilde{\Bw}(t),
\end{equation}
where $\BG_t\in\mathbb{R}^{N_{\rm r}\times T}$ is defined as 
\begin{equation}
\label{eq:G_t_def}
\BG_{t}
= \frac{1}{\sqrt{2N_{\rm t}}}
\begin{bmatrix}
\BG_{\rm R} & -\BG_{\rm I} 
\end{bmatrix}
\begin{bmatrix}
[\BA_1]_{t,:}\\
\vdots\\
[\BA_{2N_{\rm t}}]_{t,:}
\end{bmatrix}.
\end{equation}
Finally, by stacking all the observed signals $\Tilde{\By}(t)$ for all $t \in [T]$ as $\tilde{\By}_{\rm eq} = \operatorname{col}\{\tilde{\By}(1),\tilde{\By}(2),\ldots,\tilde{\By}(T) \} \in \mathbb{R}^{N_{\rm r}T \times 1}$, and similarly stacking all noise vectors $\tilde{\Bw}(t)$ for all $t \in [T]$ as $\tilde{\Bw}_{\rm eq} = \operatorname{col}\{\tilde{\Bw}(1),\tilde{\Bw}(2),\ldots,\tilde{\Bw}(T)\}$, the input-output relationship  for our RAQR-based system during all $T$ channel uses is modeled as follows: 
\begin{equation}
\label{eq:stacked_rx}
\tilde{\By}_{\rm eq} \approx \BG_{\rm eq}\Bs + \tilde{\Bw}_{\rm eq},
\end{equation}
where $\BG_{\rm eq}
=
\begin{bmatrix}
\BG_1^{\rm T} &
\cdots  &
\BG_T^{\rm T}
\end{bmatrix}^{\rm T}
\in
\mathbb{R}^{N_{\rm r}T\times T}$. From \eqref{eq:A conditions} and \eqref{eq:y_t_detection}, the equivalent channel matrix $\BG_{\rm eq}$ satisfies
\begin{equation}
\label{eq:G_eq_orthogonality}
\BG_{\rm eq}^{\rm T} \BG_{\rm eq} = \frac{1}{{2N_{\rm t}}}\|\BG\|_{\rm F}^{2} \BI_T.
\end{equation}
Accordingly, from \eqref{eq:G_eq_orthogonality}, the real information symbols are decoupled after linear combining, such that the real information $s_{i}$ is detected by $ \hat{r}_i = \frac{[\BG_{\rm eq}]_{:,i}^{\rm T} \tilde{\By}_{\rm eq}}{ \left\| [\BG_{\rm eq}]_{:,i} \right\|^2 }$. Considering $\left\| [\BG_{\rm eq}]_{:,i} \right\|^2 = \frac{1}{{2N_{\rm t}}} \|\BG\|_{\rm F}^{2} $  for all $i \in [T]$ and $[\BG_{\rm eq}]_{:,i}^{\rm T} [\BG_{\rm eq}]_{:,j} = 0$, with $i \neq j$ (cf. \eqref{eq:A conditions} and \eqref{eq:G_t_def}), $ \hat{r}_{i}$ is given by:
\begin{equation}
\label{eq: ri hat def}
        \hat{r}_{i} = s_{i} + \hat{n}_{i},
\end{equation} 
where $Z =  \|\BG\|_{\rm F}^{2}$ and $\hat{n}_{i} = \frac{2N_{\rm t}}{Z}[\BG_{\rm eq}]_{:,i}^{\rm T} \tilde{\Bw}_{\rm eq}$ is a Gaussian noise term satisfying $\hat{n}_{i} \sim \mathcal{N} (0, \hat{\sigma}^{2})$ with $\hat{\sigma}^2= \frac{N_{\rm t}\sigma^2}{Z}$. 
Hence, the real symbol $s_{i}$ is independently detected by deploying a nearest-neighbor detector as follows:\footnote{Practically speaking, in a channel-coded system, the MF output $\hat r_i$ in \eqref{eq: ri hat def} can instead be provided to a soft demapper to generate bit log-likelihood ratios for channel decoding.}
\begin{equation}
\label{eq:symbol_decision}
\hat{s}_i = \arg\min_{c\in\mathcal{S}} |\hat{r}_{i}-c|^2.
\end{equation}
In the following, we assess the proposed RAQR-OSTBC performance in terms of BEP. 
\section{Performance Analysis}
\label{appendix: proof theorrem}
To derive the analytical BEP, we first compute the symbol error probability (SEP) and then apply the Gray-coded approximation.  To this end, consider an $M$-PAM constellation comprising equally spaced points, and let $\delta$ be the difference between the adjacent constellation points. Accordingly, by considering $E_{\rm s}$ as the average energy per symbol and $E_{\rm b}$ as the average energy per bit, i.e., $E_{\rm s} = E_{\rm b} \log_{2}(M)$, we have $\delta = \sqrt{\frac{12E_{\rm b} \log_{2}(M)}{M^2 - 1}}$. Hence, one can show that for a given realization $Z=z$ of the random variable $Z$, defined after~\eqref{eq: ri hat def}, the conditional SEP, denoted by $P_{{\rm s}\mid Z}(z)$, is given by~\cite{simon2004digital}:
\begin{equation}
\label{eq: SEP z extended}
        P_{{\rm s}\mid Z}(z)  =  \frac{2\left( M-1 \right)}{M} Q\left( \sqrt{ \frac{3 z \gamma_{\rm b} \log_{2}(M)}{N_{\rm t}(M^2 - 1)} } \right),  
\end{equation}
where $\gamma_{\rm b} = E_{\rm b}/\sigma^2$ represents the average transmit signal-to-noise ratio (SNR) per bit. Next, the SEP is obtained by averaging $P_{{\rm s}\mid Z}(Z)$ over all realizations of $Z$, i.e., $\mathrm{SEP} = \mathbb{E}_{Z} \left[ P_{{\rm s}\mid Z}(Z) \right]$. Accordingly, using Craig's formula and employing \eqref{eq: SEP z extended}, the SEP takes the following form:
\begin{equation}
\label{eq: SEP def}
        \mathrm{SEP}  = 
        \frac{2\left( M-1 \right)}{M\pi} \int\nolimits_{0}^{\frac{\pi}{2}} \mathbb{E} \left[ e^{-\frac{3 Z \gamma_{\rm b} \log_{2}(M)}{2N_{\rm t}(M^2 - 1)  \sin^{2}\theta}} \right] d\theta.
\end{equation}

In order to compute \eqref{eq: SEP def}, we need to find the distribution of $Z$. To this end, by applying $a_{nkl} = \frac{\boldsymbol{\mu}^{\rm T} \boldsymbol{\epsilon}_{nkl}}{\hbar} e^{j  \phi_{nkl}}$ to \eqref{eq:multipath_channel}, the effective channel takes the form of $ [\BH]_{n,k} = \frac{1}{\sqrt{L}} \sum\nolimits_{l = 1}^{L} g_{nkl} a_{nkl}$. As stated earlier, $g_{nkl} \sim \CC \CN (0,1)$, and accordingly, it follows that $g_{nkl} a_{nkl} \sim \CC \CN (0, \left\vert a_{nkl} \right\vert^{2})$. Hence, given the sequence $\lbrace a_{nkl} \rbrace$, the channel entry $[\BH]_{n,k}$ is a complex Gaussian random variable with  $[\BH]_{n,k} \sim \CC \CN(0,\Omega_{nk})$, where  $\Omega_{nk} = \frac{1}{L} \sum_{l = 1}^{L} \left\vert a_{nkl} \right\vert^{2}$. Under the rich-scattering assumption and for sufficiently large $L$, $\Omega_{nk}$ is approximated by the common average channel power $\Omega_{\BH} $. Specifically, by the law of large numbers, we have $\Omega_{nk}       \approx  \mathbb{E} \left[ \left\vert a_{nkl} \right\vert^{2} \right] $, which results in: 
\begin{equation*}
\label{eq: Omega def}
      \Omega_{nk}      \approx  \mathbb{E} \left[ \left\vert a_{nkl} \right\vert^{2} \right]  = \frac{\boldsymbol{\mu}^{\rm T} \mathbb{E}\left[ \boldsymbol{\epsilon}_{nkl} \boldsymbol{\epsilon}_{nkl}^{\rm T} \right] \boldsymbol{\mu}}{\hbar^{2}}  \triangleq  \Omega_{\BH}.
\end{equation*}
Assuming that \(\boldsymbol{\epsilon}_{nk\ell}\) is uniformly distributed on the unit circle perpendicular to the unit incident-wave direction \(\Bu\), its conditional covariance is \(\mathbb{E}[\boldsymbol{\epsilon}_{nk\ell}\boldsymbol{\epsilon}_{nk\ell}^{\rm T}\mid\Bu]=(\BI-\Bu\Bu^{\rm T})/2\). Averaging over \(\Bu\) gives \(\mathbb{E}[\boldsymbol{\epsilon}_{nk\ell}\boldsymbol{\epsilon}_{nk\ell}^{\rm T}]=(\BI-\mathbb{E}[\Bu\Bu^{\rm T}])/2\). For \(\Bu\) uniformly distributed in the \(xy\)-plane, \(\mathbb{E}[\Bu\Bu^{\rm T}]=\frac{1}{2}\operatorname{diag}(1,1,0)\), yielding \(\Omega_{\BH}=(\mu_x^2+\mu_y^2+2\mu_z^2)/(4\hbar^2)\). Since the diagonal phase compensation in \eqref{eq:effective_channel} preserves the i.i.d. channel distribution, under the rich-scattering approximation, the entries of $\BG$ are modeled as i.i.d. random variables satisfying $[\BG]_{n,k}\sim\mathcal{CN}(0,\Omega_{\BH})$.
Hence, \(\lvert[\BG]_{n,k}\rvert^2\) are i.i.d. exponential random variables with mean \(\Omega_{\BH}\). Therefore, \(Z=\|\BG\|_{\rm F}^2=\sum_{n=1}^{N_{\rm r}}\sum_{k=1}^{N_{\rm t}}\lvert[\BG]_{n,k}\rvert^2\) follows a Gamma distribution with shape parameter \(N_{\rm t}N_{\rm r}\) and scale parameter \(\Omega_{\BH}\).

Now,  considering that $\mathbb{E} \left[ e^{-\frac{3 Z \gamma_{\rm b} \log_{2}(M)}{2N_{\rm t}(M^2 - 1)  \sin^{2}\theta}} \right]$ is the moment-generating function evaluated at a negative argument, by adopting \eqref{eq: SEP def}, SEP is obtained as follows: 
\begin{equation}
\label{eq: SEP final}
         \mathrm{SEP}  = \frac{2  \left( M-1 \right)}{M\pi} \int\limits_{0}^{\frac{\pi}{2}}  \left( 1 + \frac{3 \gamma_{\rm b} \log_{2}(M) \Omega_{\BH}}{2N_{\rm t}(M^2 - 1)  \sin^{2}\theta} \right)^{-N_{\rm t}N_{\rm r}}  d\theta.
\end{equation}
Finally, by using the Gray-coded approximation, the BEP is obtained as $\mathrm{BEP} \approx \frac{\mathrm{SEP}}{\log_{2}(M)}$. To characterize the diversity and coding gains, we consider the high-SNR regime, i.e., $\gamma_{\rm b}\to\infty$. From \eqref{eq: SEP final}, the BEP can be asymptotically approximated as $ {\rm BEP} \approx  \SfC^{-\Sfd}\gamma_{\rm b}^{-\Sfd}$, where $\Sfd = N_{\rm t}N_{\rm r}$ is the diversity order and
\(
    \SfC  = \frac{3\Omega_{\rm H}\log_2(M)}{2N_{\rm t}(M^2-1)} \left[ \frac{M-1}{M\log_2(M)} \frac{\binom{2\Sfd}{\Sfd}}{4^{\Sfd}}\right]^{-\frac{1}{\Sfd}}
\)
is the coding gain.

\section{Numerical Results}
\label{sec:results}
\begin{figure}[t]
    \centering
    \resizebox{0.82\columnwidth}{!}{%
    \input{simulationvstheory}
}
    \caption{Analytical BEP and simulated BER performance of the proposed scheme for $\eta=1$ bpcu and different values of $N_{\rm t}$.}
    \label{fig:ber_validation}
\end{figure}
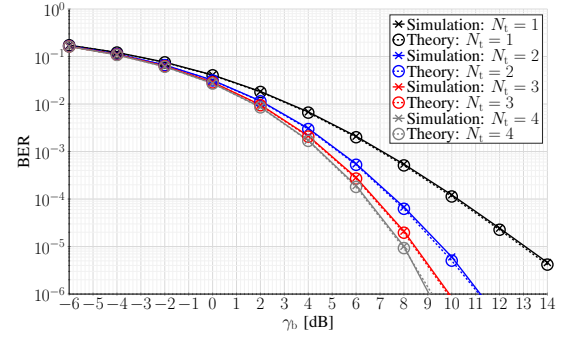

In this section, we evaluate the BER performance of the proposed RAQR-OSTBC scheme for a MIMO system with $N_{\rm r}=4$ vapor cells. The multipath channel follows the setup in~\cite{cui2025atomicmimo}, with 23 clusters, 20 paths per cluster, complex Gaussian path gains following $\CC \CN(0,1)$, uniformly distributed incident angles $\CU(-90^{\circ},90^{\circ})$, and a maximum angular spread of $5^\circ$. The Rydberg states $52D_{5/2}$ and $53P_{3/2}$ are considered for sub-6 GHz signal reception at $f_{\rm c}=5$ GHz, with transition dipole-moment vector $\boldsymbol{\mu}= [0,1785.916qa_{0},0]^{\rm T}$, where $q$ is the electron charge and $a_{0}=5.292\times10^{-11}$ m is the Bohr radius. The polarization vectors $\boldsymbol{\epsilon}_{n,k,\ell}$ and $\boldsymbol{\epsilon}_{{\rm b},m}$ are randomly drawn from the unit circle perpendicular to their corresponding incident directions. Fig.~\ref{fig:ber_validation} compares the analytical BEP and simulated BER results for $\eta=1$ bpcu and $N_{\rm t}= \{ 1,2,3,4 \}$. The close match between theory and simulation validates the derived BEP expression and the equivalent real-valued RAQR-OSTBC model. Moreover, the BER curves become steeper as $N_{\rm t}$ increases, confirming the transmit-diversity gain predicted by the analysis.

We next compare RAQR-OSTBC with SMUX and PRSS~\cite{liu2026prss} at equal spectral efficiency $\eta$. SMUX transmits independent complex symbols from all antennas and applies joint maximum-likelihood (ML) detection to the real-valued model in \eqref{eq:linear_system_modified}, yielding diversity order $N_{\rm r}/2$. For $N_{\rm t}=1$, it reduces to a single-input multi-output (SIMO) model. PRSS instead sends each complex transmit vector and its $\pi/2$-rotated replica over two channel uses, reconstructing a complex-valued linear observation for vector-level ML detection. 
Fig.~\ref{fig:comparison_baselines}(a) compares the schemes at $\eta=1$ bpcu for $N_{\rm t}\in \{1,2 \}$; SMUX with $N_{\rm t}=2$ is omitted because it would require a non-integer constellation order. For $N_{\rm t}=1$, RAQR-OSTBC and PRSS nearly coincide because transmit diversity is unavailable and both exploit the $N_{\rm r}$ receive branches. The reconstructed complex observation also explains the PRSS gain over  SIMO, whose vapor cells provide only one real projection. For $N_{\rm t}=2$, RAQR-OSTBC spreads each real symbol orthogonally across antennas and time, allowing it to experience all $N_{\rm t}N_{\rm r}$ transmit--receive links. The resulting full diversity produces the steeper high-SNR BER slope predicted by the analysis in Section~\ref{appendix: proof theorrem}.

\begin{figure*}[t]
    \centering
    \begin{minipage}{0.48\textwidth}
        \centering
        \resizebox{.75\linewidth}{!}
        {\input{eta1}}
        \centerline{(a) $\eta=1$ bpcu}
    \end{minipage}
    \hfill
    \begin{minipage}{0.48\textwidth}
        \centering
        \resizebox{.75\columnwidth}{!}
        {\input{eta2}}
        \centerline{(b) $\eta=2$ bpcu}
    \end{minipage}
    \caption{BER comparison of the proposed RAQR-OSTBC scheme with SIMO, SMUX, and PRSS for the same spectral efficiency.}
    \label{fig:comparison_baselines}
\end{figure*}

Fig.~\ref{fig:comparison_baselines}(b) illustrates the coding--diversity gain tradeoff at $\eta=2$ bpcu. SIMO employs QPSK,
whereas SMUX with $N_{\rm t}=2$ transmits BPSK from each antenna. Under the adopted i.i.d. channel and total-power normalization, their real-valued detection models have the same statistical structure, explaining the nearly overlapping BER curves. PRSS outperforms RAQR-OSTBC at low SNR because the latter requires a higher-order real constellation, resulting in a smaller minimum distance and lower coding gain. This modulation-dependent loss mainly shifts the BER curve horizontally without changing its asymptotic diversity slope. At high SNR,  the full $N_{\rm t}N_{\rm r}$ diversity of RAQR-OSTBC dominates, producing a steeper BER decay. The curve crossings therefore mark the transition from coding-gain dominance at low SNR to diversity-gain dominance at high SNR. Overall, SMUX and PRSS require joint vector processing, with PRSS additionally using a second channel use to reconstruct a complex-valued observation. In contrast, RAQR-OSTBC trades spectral efficiency for full transmit--receive diversity and low-complexity symbol-wise detection.

\section{Conclusion}
\label{sec:conclusion}
This letter has proposed a low-complexity space-time-coded framework for RAQR-assisted MIMO systems. At the transmitter, real information symbols have been encoded into space-time codewords using real orthogonal designs, enabling transmit diversity over multiple channel uses. At the receiver, a heterodyne RAQR architecture with a strong reference signal has been employed to obtain a real-valued linear observation model, which allows MF-based symbol-wise detection. The derived analytical BEP showed that the proposed scheme achieves full transmit-receive diversity and improves error performance compared with spatial-multiplexing-based benchmarks. Future work may extend the proposed framework to imperfect reference-signal conditions, neural-network receivers tailored to the nonlinear RAQR input-output mapping,  multi-user transmission, as well as joint space-time coding and precoding for scalable RAQR-enabled wireless systems.

\bibliography{References}
\bibliographystyle{IEEEtran}

\end{document}

%% file: MyCommands.tex
\newcommand{\CC}[0]{{\mathcal{C}}}

\newcommand{\CN}[0]{{\mathcal{N}}}

\newcommand{\CU}[0]{{\mathcal{U}}}

\newcommand{\Bb}[0]{{\mathbf{b}}}

\newcommand{\Bg}[0]{{\mathbf{g}}}

\newcommand{\Bs}[0]{{\mathbf{s}}}

\newcommand{\Bu}[0]{{\mathbf{u}}}
\newcommand{\Bv}[0]{{\mathbf{v}}}
\newcommand{\Bw}[0]{{\mathbf{w}}}
\newcommand{\Bx}[0]{{\mathbf{x}}}
\newcommand{\By}[0]{{\mathbf{y}}}

\newcommand{\BA}[0]{{\mathbf{A}}}

\newcommand{\BC}[0]{{\mathbf{C}}}

\newcommand{\BG}[0]{{\mathbf{G}}}
\newcommand{\BH}[0]{{\mathbf{H}}}
\newcommand{\BI}[0]{{\mathbf{I}}}

\newcommand{\BM}[0]{{\mathbf{M}}}

\newcommand{\BX}[0]{{\mathbf{X}}}

\newcommand{\Sfd}[0]{{\mathsf{d}}}

\newcommand{\SfC}[0]{{\mathsf{C}}}

\renewcommand{\Re}{\mbox{Re}}
\renewcommand{\Im}{\mbox{Im}}

\ExplSyntaxOn

\NewDocumentCommand \vect { s o m }
 {
  \IfBooleanTF {#1}
   { \vectaux*{#3} }
   { \IfValueTF {#2} { \vectaux[#2]{#3} } { \vectaux{#3} } }
 }
\DeclarePairedDelimiterX \vectaux [1] {\lbrack} {\rbrack}
 { \, \dbacc_vect:n { #1 } \, }
\cs_new_protected:Npn \dbacc_vect:n #1
 {
  \seq_set_split:Nnn \l_tmpa_seq { , } { #1 }
  \seq_use:Nn \l_tmpa_seq { \enspace }
 }
\ExplSyntaxOff

%% file: simulationvstheory.tex
\pgfplotstableread[row sep=\\,col sep=&]{
snr & BER\\
-10	&	0.27268961	\\
-8	&	0.2249013	\\
-6	&	0.1732759	\\
-4	&	0.12188911	\\
-2	&	0.07606735000000001	\\
0	&	0.04086991	\\
2	&	0.0183806	\\
4	&	0.00682655	\\
6	&	0.00209453	\\
8	&	0.00054255	\\
10	&	0.00012238	\\
12	&	2.437e-05	\\
14	&	4.61e-06	\\
16	&	7.2e-07	\\
18	&	1.5e-07	\\
20	&	2e-08	\\
22	&	0	\\
24	&	0	\\
26	&	0	\\
28	&	0	\\
30	&	0	\\
32	&	0	\\
34	&	0	\\
}\SIMone

\pgfplotstableread[row sep=\\,col sep=&]{
snr & BER\\
-10	&	0.2723686504022456	\\
-8	&	0.2244436985573527	\\
-6	&	0.1727569483208488	\\
-4	&	0.1212629908379874	\\
-2	&	0.07540040558684832	\\
0	&	0.04025811897863134	\\
2	&	0.01796442303478789	\\
4	&	0.006599448532939711	\\
6	&	0.002001189953931056	\\
8	&	0.0005110302130031446	\\
10	&	0.0001133583726240019	\\
12	&	2.259214872145835e-05	\\
14	&	4.169591750239887e-06	\\
16	&	7.299299217410077e-07	\\
18	&	1.233520059090354e-07	\\
20	&	2.036959164337463e-08	\\
22	&	3.313904007878578e-09	\\
24	&	5.340083458618923e-10	\\
26	&	8.55289084471725e-11	\\
28	&	1.364582930513994e-11	\\
30	&	2.171821920252136e-12	\\
32	&	3.451249299220076e-13	\\
34	&	5.479030264849082e-14	\\
}\THRone

\pgfplotstableread[row sep=\\,col sep=&]{
snr & BER\\
-10	&	0.26832185	\\
-8	&	0.21887765	\\
-6	&	0.1658658	\\
-4	&	0.1128965	\\
-2	&	0.06618865	\\
0	&	0.0317187	\\
2	&	0.01164695	\\
4	&	0.0030798	\\
6	&	0.0005496500000000001	\\
8	&	6.774999999999999e-05	\\
10	&	6.05e-06	\\
12	&	3e-07	\\
14	&	0	\\
16	&	0	\\
18	&	0	\\
20	&	0	\\
22	&	0	\\
24	&	0	\\
26	&	0	\\
28	&	0	\\
30	&	0	\\
32	&	0	\\
34	&	0	\\
}\SIMtwo

\pgfplotstableread[row sep=\\,col sep=&]{
snr & BER\\
-10	&	0.2680117948509358	\\
-8	&	0.2187811779151995	\\
-6	&	0.165541097669591	\\
-4	&	0.1125328394165561	\\
-2	&	0.06585030724007748	\\
0	&	0.03138598175730168	\\
2	&	0.01141917563037535	\\
4	&	0.002971588093006267	\\
6	&	0.0005263960595217835	\\
8	&	6.237250126477072e-05	\\
10	&	5.050937745862616e-06	\\
12	&	2.948551895484723e-07	\\
14	&	1.328833177506144e-08	\\
16	&	4.946531230977876e-10	\\
18	&	1.609973875769916e-11	\\
20	&	4.784516812745054e-13	\\
22	&	1.338922189862571e-14	\\
24	&	3.603338094166743e-16	\\
26	&	9.456769282251694e-18	\\
28	&	2.442370688680644e-19	\\
30	&	6.243803724181268e-21	\\
32	&	1.585905823744483e-22	\\
34	&	4.011690227230496e-24	\\
}\THRtwo

\pgfplotstableread[row sep=\\,col sep=&]{
snr & BER\\
-10	&	0.26677305	\\
-8	&	0.217026125	\\
-6	&	0.16321505	\\
-4	&	0.109888975	\\
-2	&	0.0629184	\\
0	&	0.028699925	\\
2	&	0.009623425	\\
4	&	0.0021319	\\
6	&	0.000284025	\\
8	&	2.1075e-05	\\
10	&	8.5e-07	\\
12	&	0	\\
14	&	0	\\
16	&	0	\\
18	&	0	\\
20	&	0	\\
22	&	0	\\
24	&	0	\\
26	&	0	\\
28	&	0	\\
30	&	0	\\
32	&	0	\\
34	&	0	\\
}\SIMthree

\pgfplotstableread[row sep=\\,col sep=&]{
snr & BER\\
-10	&	0.2665344987484498	\\
-8	&	0.2168565495424954	\\
-6	&	0.1630813217410372	\\
-4	&	0.1095529770745425	\\
-2	&	0.06261314655186268	\\
0	&	0.02846992496829583	\\
2	&	0.009443712565624814	\\
4	&	0.002065875001388231	\\
6	&	0.0002698370753021121	\\
8	&	1.95693369427554e-05	\\
10	&	7.712195862032242e-07	\\
12	&	1.716028096596296e-08	\\
14	&	2.346468131093816e-10	\\
16	&	2.190233582045849e-12	\\
18	&	1.544657843409626e-14	\\
20	&	8.955347076499937e-17	\\
22	&	4.548655045728181e-19	\\
24	&	2.117401157211531e-21	\\
26	&	9.313974396345389e-24	\\
28	&	3.950655184664175e-26	\\
30	&	1.637268505191456e-28	\\
32	&	6.685930695162749e-31	\\
34	&	2.704838772318899e-33	\\
}\THRthree

\pgfplotstableread[row sep=\\,col sep=&]{
snr & BER\\
-10	&	0.265817	\\
-8	&	0.216222875	\\
-6	&	0.16212475	\\
-4	&	0.108270375	\\
-2	&	0.061214375	\\
0	&	0.027256375	\\
2	&	0.008646875	\\
4	&	0.00172325	\\
6	&	0.00019175	\\
8	&	1.0125e-05	\\
10	&	1.25e-07	\\
12	&	0	\\
14	&	0	\\
16	&	0	\\
18	&	0	\\
20	&	0	\\
22	&	0	\\
24	&	0	\\
26	&	0	\\
28	&	0	\\
30	&	0	\\
32	&	0	\\
34	&	0	\\
}\SIMfour

\pgfplotstableread[row sep=\\,col sep=&]{
snr & BER\\
-10	&	0.2657913659880685	\\
-8	&	0.2158874683676607	\\
-6	&	0.1618413383868457	\\
-4	&	0.1080500626037568	\\
-2	&	0.06098544239990608	\\
0	&	0.02702409272719525	\\
2	&	0.008503760998586306	\\
4	&	0.001675529471983721	\\
6	&	0.0001798407180927052	\\
8	&	9.272438682225893e-06	\\
10	&	2.127896019702483e-07	\\
12	&	2.174585632040673e-09	\\
14	&	1.06962262105231e-11	\\
16	&	2.878781513495883e-14	\\
18	&	4.888795389129686e-17	\\
20	&	5.96070966347265e-20	\\
22	&	5.784303949270941e-23	\\
24	&	4.818690800174674e-26	\\
26	&	3.632419258219913e-29	\\
28	&	2.566886992212213e-32	\\
30	&	1.740372113680606e-35	\\
32	&	1.14927740814541e-38	\\
34	&	7.46340170163336e-42	\\
}\THRfour

\begin{tikzpicture}[scale=.33]

\begin{axis}[%
width=2.9\columnwidth,
height=1.8\columnwidth,
axis lines = left,
xmin=-6,
xlabel={\Huge $\gamma_{\rm b}$ [dB]},
ymode=log,
ymin=10^-6,
ymax=10^0,
ylabel={\Huge BER},
ylabel near ticks,
    grid=both,
    major grid style={line width=.2pt,draw=gray!30},
    grid style={line width=.1pt, draw=gray!10},
    minor tick num=5,
    legend pos = north east,
legend style={at={(0.99,.98)},legend cell align=left, align=left, draw=white!15!black},
ticklabel style={font=\Huge},
]
\addplot[mark=x, line width=1pt, mark size=7pt,mark options={solid},draw=black, line width=2pt] table[x=snr,y=BER]{\SIMone};
\addlegendentry{\Huge Simulation: $N_{\rm t} = 1$}

\addplot[mark=o, line width=1pt, mark size=8pt,mark options={solid},draw=black, dashed,line width=2pt] table[x=snr,y=BER]{\THRone};
\addlegendentry{\Huge Theory: $N_{\rm t} = 1$}

\addplot[mark=x, line width=1pt, mark size=7pt,mark options={solid},draw=blue, line width=2pt] table[x=snr,y=BER]{\SIMtwo};
\addlegendentry{\Huge Simulation: $N_{\rm t} = 2$}

\addplot[mark=o, line width=1pt, mark size=8pt,mark options={solid},draw=blue, dashed,line width=2pt] table[x=snr,y=BER]{\THRtwo};
\addlegendentry{\Huge Theory: $N_{\rm t} = 2$}

\addplot[mark=x, line width=1pt, mark size=7pt,mark options={solid},draw=red, line width=2pt] table[x=snr,y=BER]{\SIMthree};
\addlegendentry{\Huge Simulation: $N_{\rm t} = 3$}

\addplot[mark=o, line width=1pt, mark size=8pt,mark options={solid},draw=red, dashed,line width=2pt] table[x=snr,y=BER]{\THRthree};
\addlegendentry{\Huge Theory: $N_{\rm t} = 3$}

\addplot[mark=x, line width=1pt, mark size=7pt,mark options={solid},draw=gray, line width=2pt] table[x=snr,y=BER]{\SIMfour};
\addlegendentry{\Huge Simulation: $N_{\rm t} = 4$}

\addplot[mark=o, line width=1pt, mark size=8pt,mark options={solid},draw=gray, dashed,line width=2pt] table[x=snr,y=BER]{\THRfour};
\addlegendentry{\Huge Theory: $N_{\rm t} = 4$}

\end{axis}
\end{tikzpicture}%

%% file: eta1.tex
\pgfplotstableread[row sep=\\,col sep=&]{
snr & BER\\
-10	&	0.27268961	\\
-8	&	0.2249013	\\
-6	&	0.1732759	\\
-4	&	0.12188911	\\
-2	&	0.07606735000000001	\\
0	&	0.04086991	\\
2	&	0.0183806	\\
4	&	0.00682655	\\
6	&	0.00209453	\\
8	&	0.00054255	\\
10	&	0.00012238	\\
12	&	2.437e-05	\\
14	&	4.61e-06	\\
16	&	7.2e-07	\\
18	&	1.5e-07	\\
20	&	2e-08	\\
22	&	0	\\
24	&	0	\\
26	&	0	\\
28	&	0	\\
30	&	0	\\
32	&	0	\\
34	&	0	\\
}\STBCone

\pgfplotstableread[row sep=\\,col sep=&]{
snr & BER\\
-10	&	0.26832185	\\
-8	&	0.21887765	\\
-6	&	0.1658658	\\
-4	&	0.1128965	\\
-2	&	0.06618865	\\
0	&	0.0317187	\\
2	&	0.01164695	\\
4	&	0.0030798	\\
6	&	0.0005496500000000001	\\
8	&	6.774999999999999e-05	\\
10	&	6.05e-06	\\
12	&	3e-07	\\
14	&	0	\\
16	&	0	\\
18	&	0	\\
20	&	0	\\
22	&	0	\\
24	&	0	\\
26	&	0	\\
28	&	0	\\
30	&	0	\\
32	&	0	\\
34	&	0	\\
}\STBCtwo

\pgfplotstableread[row sep=\\,col sep=&]{
snr & BER\\
-10	&	0.27264705	\\
-8	&	0.22476015	\\
-6	&	0.17327315	\\
-4	&	0.1218679	\\
-2	&	0.07606300000000001	\\
0	&	0.04076825	\\
2	&	0.0184261	\\
4	&	0.00679475	\\
6	&	0.00208845	\\
8	&	0.0005442999999999999	\\
10	&	0.00012115	\\
12	&	2.685e-05	\\
14	&	4.4e-06	\\
16	&	5e-07	\\
18	&	2e-07	\\
20	&	0	\\
22	&	0	\\
24	&	0	\\
26	&	0	\\
28	&	0	\\
30	&	0	\\
32	&	0	\\
34	&	0	\\
}\PRSSone

\pgfplotstableread[row sep=\\,col sep=&]{
snr & BER\\
-10	&	0.27742574	\\
-8	&	0.23145053	\\
-6	&	0.18123361	\\
-4	&	0.13002193	\\
-2	&	0.08274138	\\
0	&	0.04511886	\\
2	&	0.02047058	\\
4	&	0.00758373	\\
6	&	0.00232113	\\
8	&	0.00059535	\\
10	&	0.00013054	\\
12	&	2.566e-05	\\
14	&	4.59e-06	\\
16	&	9e-07	\\
18	&	1.4e-07	\\
20	&	3e-08	\\
22	&	1e-08	\\
24	&	0	\\
26	&	0	\\
28	&	0	\\
30	&	0	\\
32	&	0	\\
34	&	0	\\
}\PRSStwo


\pgfplotstableread[row sep=\\,col sep=&]{
snr & BER\\
-10	&	0.28108394	\\
-8	&	0.23576904	\\
-6	&	0.18707796	\\
-4	&	0.13855325	\\
-2	&	0.09437795	\\
0	&	0.05874153	\\
2	&	0.03325805	\\
4	&	0.01729823	\\
6	&	0.0083351	\\
8	&	0.00378956	\\
10	&	0.00165254	\\
12	&	0.00069411	\\
14	&	0.00028679	\\
16	&	0.00011834	\\
18	&	4.802e-05	\\
20	&	1.93e-05	\\
22	&	7.869999999999999e-06	\\
24	&	3.14e-06	\\
26	&	1.35e-06	\\
28	&	5.2e-07	\\
30	&	2.2e-07	\\
32	&	8e-08	\\
34	&	3e-08	\\
}\SMUXone

\begin{tikzpicture}[scale=.25]

\begin{axis}[%
width=2.9\columnwidth,
height=2\columnwidth,
axis lines = left,
xmin=-6,
xlabel={\Huge $\gamma_{\rm b}$ [dB]},
ymode=log,
ymin=10^-6,
ymax=10^0,
ylabel={\Huge BER},
ylabel near ticks,
    grid=both,
    major grid style={line width=.2pt,draw=gray!30},
    grid style={line width=.1pt, draw=gray!10},
    minor tick num=5,
    legend pos = north east,
legend style={at={(0.99,.98)},legend cell align=left, align=left, draw=white!15!black},
ticklabel style={font=\Huge},
]
\addplot[mark=square, line width=1pt, mark size=5pt,mark options={solid},draw=black, dashed,line width=2pt] table[x=snr,y=BER]{\STBCone};
\addlegendentry{\Huge  RAQR-OSTBC: $N_{\rm t} = 1$}

\addplot[mark=square, line width=1pt, mark size=5pt,mark options={solid},draw=blue, dashed,line width=2pt] table[x=snr,y=BER]{\STBCtwo};
\addlegendentry{\Huge  RAQR-OSTBC: $N_{\rm t} = 2$}

\addplot[mark=+, line width=1pt, mark size=5pt,mark options={solid},draw=red,line width=2pt] table[x=snr,y=BER]{\PRSSone};
\addlegendentry{\Huge PRSS: $N_{\rm t} = 1$}

\addplot[mark=+, line width=1pt, mark size=5pt,mark options={solid},draw=green,line width=2pt] table[x=snr,y=BER]{\PRSStwo};
\addlegendentry{\Huge PRSS: $N_{\rm t} = 2$}

\addplot[mark=o, line width=1pt, mark size=5pt,mark options={solid},solid,draw=orange,line width=2pt] table[x=snr,y=BER]{\SMUXone};
\addlegendentry{\Huge SIMO}


\end{axis}
\end{tikzpicture}%

%% file: eta2.tex
\pgfplotstableread[row sep=\\,col sep=&]{
snr & BER\\
-10	&	0.3187267	\\
-8	&	0.2739622	\\
-6	&	0.22623905	\\
-4	&	0.17871745	\\
-2	&	0.13368535	\\
0	&	0.09226535	\\
2	&	0.0570528	\\
4	&	0.0304985	\\
6	&	0.01368985	\\
8	&	0.00504905	\\
10	&	0.00154255	\\
12	&	0.00040185	\\
14	&	0.0000883	\\
16	&	0.0000165	\\
18	&	0.0000038	\\
20	&	0.0000007	\\
22	&	0.0000002	\\
24	&	0	\\
26	&	0	\\
28	&	0	\\
30	&	0	\\
32	&	0	\\
34	&	0	\\
}\STBCone

\pgfplotstableread[row sep=\\,col sep=&]{
snr & BER\\
-10	&	0.314612375	\\
-8	&	0.26791865	\\
-6	&	0.219042125	\\
-4	&	0.17111965	\\
-2	&	0.126019475	\\
0	&	0.0846413	\\
2	&	0.04935825	\\
4	&	0.023573175	\\
6	&	0.008606825	\\
8	&	0.002254525	\\
10	&	0.000408275	\\
12	&	0.000050325	\\
14	&	0.000004575	\\
16	&	0.00000015	\\
18	&	0.000000025	\\
20	&	0	\\
22	&	0	\\
24	&	0	\\
26	&	0	\\
28	&	0	\\
30	&	0	\\
32	&	0	\\
34	&	0	\\
}\STBCtwo

\pgfplotstableread[row sep=\\,col sep=&]{
snr & BER\\
-10	&	0.31883625	\\
-8	&	0.27387075	\\
-6	&	0.22625515	\\
-4	&	0.17893875	\\
-2	&	0.1336572	\\
0	&	0.0924756	\\
2	&	0.05706955	\\
4	&	0.03045625	\\
6	&	0.01367875	\\
8	&	0.00507335	\\
10	&	0.00155225	\\
12	&	0.0003939	\\
14	&	0.0000918	\\
16	&	0.0000182	\\
18	&	0.0000034	\\
20	&	0.00000065	\\
22	&	0.0000001	\\
24	&	0	\\
26	&	0	\\
28	&	0	\\
30	&	0	\\
32	&	0	\\
34	&	0	\\
}\PRSSone

\pgfplotstableread[row sep=\\,col sep=&]{
snr & BER\\
-10	&	0.282446	\\
-8	&	0.238444	\\
-6	&	0.19009575	\\
-4	&	0.13960275	\\
-2	&	0.0909575	\\
0	&	0.05074325	\\
2	&	0.023365	\\
4	&	0.00877475	\\
6	&	0.002675	\\
8	&	0.00066325	\\
10	&	0.00014225	\\
12	&	0.000031	\\
14	&	0.00000425	\\
16	&	0.0000005	\\
18	&	0	\\
20	&	0	\\
22	&	0	\\
24	&	0	\\
26	&	0	\\
28	&	0	\\
30	&	0	\\
32	&	0	\\
34	&	0	\\
}\PRSStwo

\pgfplotstableread[row sep=\\,col sep=&]{
snr & BER\\
-10	&	0.290036735	\\
-8	&	0.248052765	\\
-6	&	0.20212472	\\
-4	&	0.15455807	\\
-2	&	0.10906964	\\
0	&	0.07004753	\\
2	&	0.040707345	\\
4	&	0.021499315	\\
6	&	0.010464635	\\
8	&	0.00476578	\\
10	&	0.002071365	\\
12	&	0.000874655	\\
14	&	0.000358155	\\
16	&	0.0001468	\\
18	&	0.000059085	\\
20	&	0.00002371	\\
22	&	0.00000926	\\
24	&	0.0000038	\\
26	&	0.00000161	\\
28	&	0.000000595	\\
30	&	0.00000023	\\
32	&	0.0000001	\\
34	&	0.00000004	\\
}\SMUXtwo

\pgfplotstableread[row sep=\\,col sep=&]{
snr & BER\\
-10	&	0.3259941	\\
-8	&	0.28405951	\\
-6	&	0.23907029	\\
-4	&	0.19334206	\\
-2	&	0.14904505	\\
0	&	0.10802811	\\
2	&	0.07262869	\\
4	&	0.04478527	\\
6	&	0.02522436	\\
8	&	0.01302264	\\
10	&	0.00628048	\\
12	&	0.0028465	\\
14	&	0.00124076	\\
16	&	0.00052697	\\
18	&	0.00021496	\\
20	&	0.00008699	\\
22	&	0.00003706	\\
24	&	0.00001486	\\
26	&	0.00000595	\\
28	&	0.00000234	\\
30	&	0.00000087	\\
32	&	0.0000004	\\
34	&	0.00000014	\\
}\SMUXone

\begin{tikzpicture}[scale=.25]

\begin{axis}[%
width=2.9\columnwidth,
height=2\columnwidth,
axis lines = left,
xmin=-6,
xlabel={\Huge $\gamma_{\rm b}$ [dB]},
ymode=log,
ymin=10^-6,
ymax=10^0,
ylabel={\Huge BER},
ylabel near ticks,
    grid=both,
    major grid style={line width=.2pt,draw=gray!30},
    grid style={line width=.1pt, draw=gray!10},
    minor tick num=5,
    legend pos = north east,
legend style={at={(0.99,.98)},legend cell align=left, align=left, draw=white!15!black},
ticklabel style={font=\Huge},
]
\addplot[mark=square, line width=1pt, mark size=5pt,mark options={solid},draw=black, dashed,line width=2pt] table[x=snr,y=BER]{\STBCone};
\addlegendentry{\Huge  RAQR-OSTBC: $N_{\rm t} = 1$}

\addplot[mark=square, line width=1pt, mark size=5pt,mark options={solid},draw=blue, dashed,line width=2pt] table[x=snr,y=BER]{\STBCtwo};
\addlegendentry{\Huge  RAQR-OSTBC: $N_{\rm t} = 2$}

\addplot[mark=+, line width=1pt, mark size=5pt,mark options={solid},draw=red,line width=2pt] table[x=snr,y=BER]{\PRSSone};
\addlegendentry{\Huge PRSS: $N_{\rm t} = 1$}

\addplot[mark=+, line width=1pt, mark size=5pt,mark options={solid},draw=green,line width=2pt] table[x=snr,y=BER]{\PRSStwo};
\addlegendentry{\Huge PRSS: $N_{\rm t} = 2$}

\addplot[mark=o, line width=1pt, mark size=8pt,mark options={solid},solid,draw=orange,line width=2pt] table[x=snr,y=BER]{\SMUXtwo};
\addlegendentry{\Huge SIMO}

\addplot[mark=o, line width=1pt, mark size=5pt,mark options={solid},dashdotted, draw=violet,line width=2pt] table[x=snr,y=BER]{\SMUXtwo};
\addlegendentry{\Huge SMUX: $N_{\rm t} = 2$}

\end{axis}
\end{tikzpicture}%